\documentclass[5p,times,final]{elsarticle}

\usepackage{lineno,hyperref}

\usepackage{stfloats}

\journal{Journal of Magnetism and Magnetic Materials}

\begin{document}

\begin{frontmatter}

\title{Magnetic and magnetoelectric properties of $A$FeF$_5$ ($A$ = Ca, Sr) spin-chain compounds}




\author[IPhys]{N. V. Ter-Oganessian\corref{mycorrespondingauthor}}
\cortext[mycorrespondingauthor]{Corresponding author}
\ead{teroganesyan@sfedu.ru}
\author[MIC,MATH]{S. A. Guda}
\author[IPhys]{V. P. Sakhnenko}
\author[Yerevan,Yerevan1]{V. Ohanyan}

\address[IPhys]{Institute of Physics, Southern Federal University, 194 Stachki pr., 344090 Rostov-on-Don, Russia}
\address[MIC]{The Smart Materials Research Institute, Southern Federal University, 178/24 Sladkova st., 344090 Rostov-on-Don, Russia}
\address[MATH]{Institute of Mathematics, Mechanics and Computer Science, Southern Federal University, Milchakova 8a, 344090 Rostov-on-Don, Russia}
\address[Yerevan]{Department of Theoretical Physics, Yerevan State University, Alex Manoogian 1, 0025 Yerevan, Armenia}
\address[Yerevan1]{CANDLE Synchrotron Research Institute, 31 Acharyan Str., 
0040 Yerevan, Armenia}




\begin{abstract}
Fluorides in general are characterized by big variety of crystal structures, whereas those containing transition metals also often show sizable magnetic properties. The tendency of fluorine to form linear chain structures in many cases results in low-dimensional magnetism. Despite the plethora of magnetic phenomena in fluorides, their magnetoelectric properties are less studied than those of oxides. In the present work we theoretically study the magnetic and magnetoelectric properties of spin-chain compounds CaFeF$_5$ and SrFeF$_5$. The density functional theory is employed for determination of magnetic exchange constants, which are then used in Monte Carlo calculations. The symmetry analysis reveals that CaFeF$_5$ does not show magnetoelectric properties, whereas SrFeF$_5$ is a multiferroic.
\end{abstract}



\end{frontmatter}

\nolinenumbers

\section{Introduction}

Multiferroics and specifically magnetoelectrics continue to be one of the focal points in condensed matter physics due to promising practical applications. Combination of magnetic and electric properties in a single material opens up new  opportunities for creating various sensors, logical elements, or spin electronic devices~\cite{Pyatakov_UFN_Review}. However, interacting magnetic and electric subsystems are also of fundamental interest. In magnetism low-dimensional systems continue to serve as playground to explore exotic quantum phenomena such as spin liquids, Bose-Einstein condensation, or spin-Peierls transition~\cite{Vasiliev_npjQM_Milestones}. Such quantum phenomena can be easily tuned, changed, or completely destroyed by small external influences. Vast experimental data on magnetoelectric compounds allows concluding that interaction of magnetic and electric subsystems is frequent for both simple or more complex magnetic orderings. Therefore, exploiting the electric subsystem as a means for tuning the magnetic subsystems via, e.g., application of external electric field, can significantly extend the plethora of magnetic phenomena in low-dimensional systems or potentially result in new effects.

Fluoride compounds offer unique opportunities for low-di\-men\-sional magnetic systems, because fluorine exhibits strong preference for linear bridging modes, which can provide the necessary topology. A great number of fluoride inorganic compounds have been synthesized and studied to date, ranging from 0D to 3D connectivity of $M$F$_x$ polyhedra surrounding the metal $M$~\cite{Hagenmuller_Book,Leblanc_ChemRev_Crystal_Chemistry}. Even when the magnetic network has three-di\-men\-sio\-nal character, the magnetic interactions are often anisotropic leading to low-di\-men\-sio\-nal magnetism.

Many fluoride compounds have already been shown to exhibit multiferroic properties and the available experimental data were summarized in recent reviews~\cite{Scott_Why_are_there,Calestani_Multiferroic_Fluorides}, however most of the compounds are the, so-called, type-I multiferroics with magnetic and electric subsystems ordering independently~\cite{Khomskii_Classifying}. Therefore, from our point of view searching for new fluoride type-II multiferroic compounds, which have stronger magnetoelectric coupling than type-I compounds, can be fruitful because (i) fluorides have diverse compositions and crystal structures, (ii) in many cases the exchange coupling in them is high, and (iii) low-dimensional magnetism in combination with magnetoelectric phenomena can potentially result in new physics.

In this work we theoretically study the magnetic and possible \hfill magnetoelectric \hfill phenomena \hfill in \hfill spin-chain \hfill compounds\\ CaFeF$_5$ and SrFeF$_5$. We find the magnetic exchange constants using the density functional theory, which are then used in Monte Carlo calculations. Supplemented with the symmetry analysis, the study reveals the magnetic and magnetoelectric properties of these fluorides.

\section{Methods}

We performed density functional theory (DFT) calculations using the Vienna {\it Ab-initio} Simulation Package (VASP)~\cite{Kresse_1996} and the projected augmented wave method~\cite{Bloechl_1994}. The generalized gradient approximation (GGA) of exchange correlation corrected by means of the GGA+U formalism for the Fe atoms with $U_{\rm eff}=U-J=3$~eV was used within the Dudarev approach~\cite{Dudarev_1998}. The energy cutoff was 500~eV, whereas the Brillouin zone integration was done using the Monkhorst-Pack scheme.

The crystal structure relaxation was done using the stopping criterion for absolute values of forces on atoms of 10$^{-3}$~eV/\AA. For\hfill the\hfill determination\hfill of\hfill exchange\hfill constants\hfill spin\hfill polarized\hfill col\-linear\hfill calculations\hfill were\hfill used\hfill and\hfill the\hfill Hamiltonian\hfill was\hfill fitted\hfill to\hfill relative\hfill total energies of different collinear magnetic structures.

The \hfill obtained \hfill exchange \hfill constants \hfill were \hfill used \hfill in \hfill classical\\ Monte Carlo (MC) simulations of the Hamiltonian
\begin{equation}
\mathcal{H}=\sum_{ij}J_{ij}\vec{S}_i\cdot\vec{S}_j+\sum_{i}\left(
D_xS_{ix}^2+
D_yS_{iy}^2+
D_zS_{iz}^2\right),\label{eq:Hamiltonian}
\end{equation}
where $\vec{S}$ are classical vectors of unit length and $D_\alpha$ ($\alpha=x,y,z$) are anisotropy constants. In the Hamiltonian Eq.~(\ref{eq:Hamiltonian}) we account only for the anisotropy terms $D_\alpha\neq0$, which we consider being the same for all spins, pursuing a minimal model to characterize the magnetic structure. The Metropolis scheme was used for MC simulations and the size of the simulation box was chosen large enough as to not influence significantly the results (typically larger than $15\times15\times15$ unit cells). After each change of temperature the system was allowed to relax for $5\cdot10^{3}$ Monte Carlo steps per spin (MCS), whereas the statistical information was subsequently gathered over the next $10^{4}$~MCS. The crystal and magnetic structures in this work were visualized using the software VESTA~\cite{VESTA}.

\section{Crystal structure and magnetic exchange interactions}

\begin{figure*}
\centering
\includegraphics[width=17.0cm]{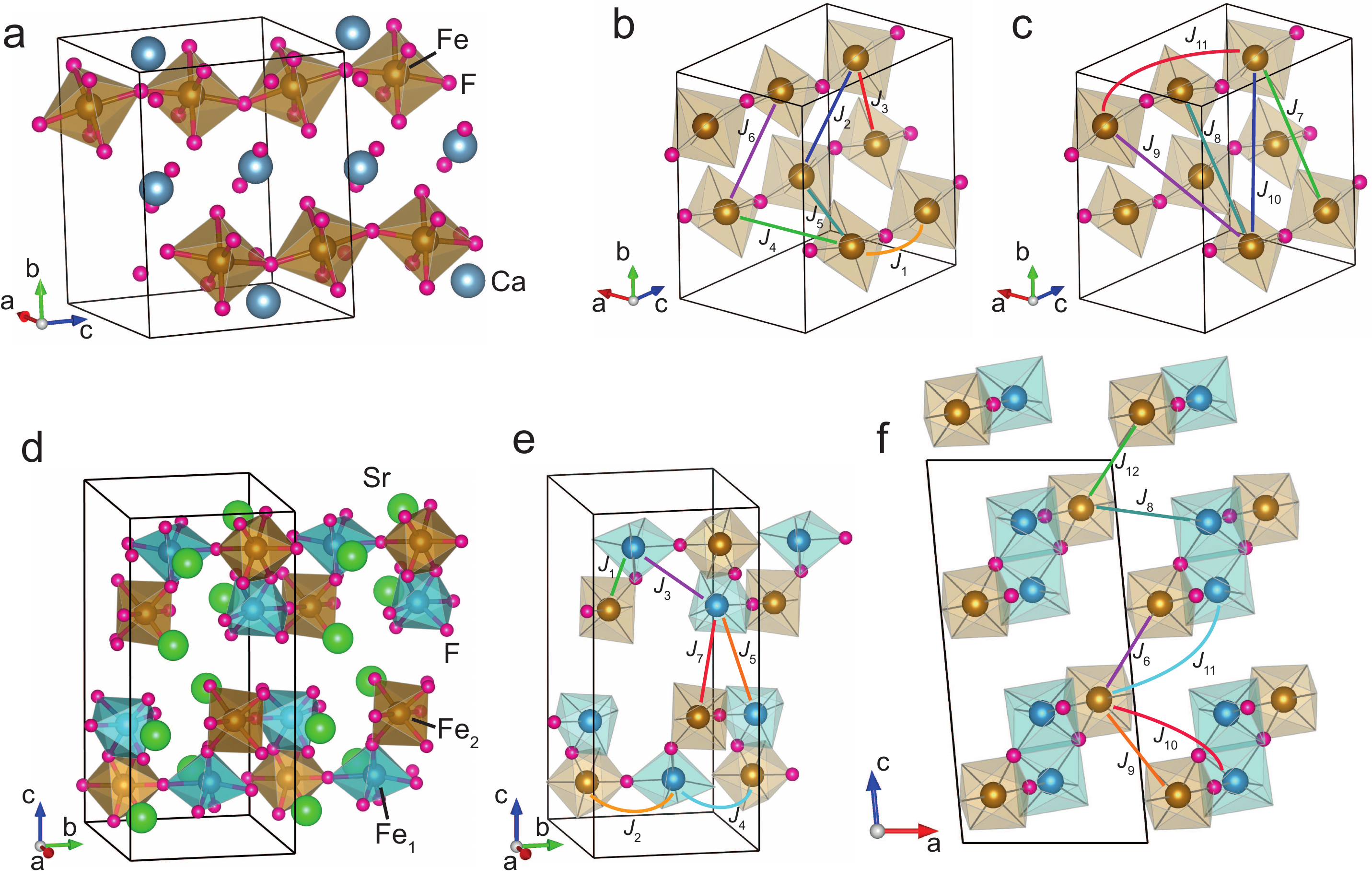}
\caption{\label{fig:CaFeF5_struct} (Color online) (a) The crystal structure of CaFeF$_5$ showing chains of octahedra FeF$_6$ along the $c$ axis. (b) and (c) Magnetic exchange paths in CaFeF$_5$. (d) The crystal structure of SrFeF$_5$ showing chains of octahedra FeF$_6$ along the $c$ axis. (e) and (f) Magnetic exchange paths in SrFeF$_5$.}
\end{figure*}

The \hfill compounds \hfill with \hfill the \hfill general \hfill chemical \hfill formula\\ $A^{2+}M^{3+}$F$_5$ ($A$ = Ca, Sr, Ba; $M$ = transition metal) present many examples of 1D chain structures, which contain isolated ($M$F$_5$)$_n$ chains of octahedra. The arrangement of infinite chains depends on the ionic radii of both $A^{2+}$ and $M^{3+}$ and several structural types have been found~\cite{Babel_CrystalChem_Hagenmueller}. CaFeF$_5$ and CaCrF$_5$ present two structural types, in which the $M$F$_6$ octahedra are linked via opposite ({\it trans-})corners and the main difference is in the positions of Ca$^{2+}$ ions. Several other fluorides, e.g., CaVF$_5$ and CaTiF$_5$, crystallize in the structure of CaCrF$_5$, whereas the structure of CaFeF$_5$ is also shared by, e.g.,  CdCrF$_5$ and CdGaF$_5$~\cite{Babel_CrystalChem_Hagenmueller}.

In contrast to the CaFeF$_5$ and CaCrF$_5$ structures, in which the chains of Fe$^{3+}$ ions are almost or completely linear (177.8$^\circ$ and 180$^\circ$, respectively), the SrFeF$_5$-type structure contains helicoidal (FeF$_5$)$_n$ chains and the octahedra are connected by corners in the {\it cis}-position~\cite{VonDerMuehll_SrFeF5}. Two octahedral chains with opposite rotations are present in the structure. To this structural type belong also SrVF$_5$, SrCoF$_5$, and SrTiF$_5$.

The crystal structures of CaFeF$_5$ and SrFeF$_5$ are shown in Figs.~\ref{fig:CaFeF5_struct}(a,d). Tables~\ref{tab:CaFeF5_structure} and~\ref{tab:SrFeF5_structure} present the lattice parameters and atomic positions obtained by DFT calculations in comparison with the experimental results taken from literature. Overall we obtain good correspondence with the experimental results with lattice parameters difference of not more than 1.5\%~\cite{VonDerMuehll_SrFeF5,Graulich_1998}.

We restrict the DFT calculations of magnetic exchange constants $J_i$ of CaFeF$_5$ to eleventh nearest neighbor (NN). The exchange paths in CaFeF$_5$ are shown in Figs.~\ref{fig:CaFeF5_struct}(b,c). Thus, the largest distance between the Fe$^{3+}$ ions, for which the exchange constant is determined corresponds to next nearest neighbors (NNN) within the chains. The obtained values of $J_i$ ordered with respect to increasing distance between the spins are given in Table~\ref{tab:Js_CaFeF5}. It can be found that $J_1$ and $J_{11}$ are the only intrachain interactions representing NN and NNN exchanges within the chains, respectively, whereas all other interactions are between different chains. The dominant exchange interaction is NN within a chain ($J_1$), while the NNN interaction $J_{11}$ is almost two orders of magnitude smaller. The strongest interchain interaction $J_7$ is roughly 30 times smaller than $J_1$. Moreover, given a strong tendency for $\ldots$-up-down-up-down-$\ldots$ ordering within the chains, some interchain interactions introduce frustration in the system, e.g., the interactions $J_7$ and $J_9$ cannot be simultaneously satisfied. Overall, CaFeF$_5$ can be characterized as having a strong NN antiferromagnetic intrachain interaction, weak NNN antiferromagnetic exchange within the chains, and weak interchain magnetic coupling.

The \hfill magnetic \hfill exchange \hfill paths \hfill of \hfill SrFeF$_5$, \hfill shown \hfill in\\ Figs.~\ref{fig:CaFeF5_struct}(e,f), are determined up to the twelfth nearest neighbor. The calculated values of $J_i$ are given in Table~\ref{tab:Js_SrFeF5}. It can be found that the chains are composed of alternating Fe$_1$ and Fe$_2$ octahedra with alternating NN exchange constants $J_1$ and $J_2$, whose values are comparable to that in CaFeF$_5$, whereas $J_2/J_1\approx1.38$. The NNN intrachain interaction is $J_3$, whereas interaction $J_4$ can be considered as next next nearest neighbor along the chain. The exchange interactions $J_i$ with $i$ = 5 -- 12 are interchain couplings. Similar to the case of CaFeF$_5$ the dominant exchange interactions in SrFeF$_5$ are NN intrachain couplings, whereas further neighbor exchanges within chains are significantly smaller. However, in contrast to CaFeF$_5$, interchain interactions in SrFeF$_5$ are relatively stronger.

\begin{table*}
\caption{Calculated magnetic exchange constants for CaFeF$_5$ and respective Fe -- Fe distances.\label{tab:Js_CaFeF5}}
\centering
\begin{tabular}{ccccccc}
\hline
 & $J_1$ & $J_2$ & $J_3$ & $J_4$ & $J_5$ & $J_6$ \\
\hline
Fe -- Fe, \AA & 3.800 & 5.428 & 5.484 & 5.492 & 5.506 & 5.559 \\
J, meV &  21.9 & -0.037 & 0.076 & 0.18 & 0.086 & 0.24 \\
\hline
\hline
& $J_7$ & $J_8$ & $J_9$ & $J_{10}$ & $J_{11}$ &  \\
\hline
Fe -- Fe, \AA & 6.437 & 6.549 & 7.506 & 7.540 & 7.599 &  \\
J, meV &  0.78 & -0.2 & 0.14 & 0.01 & 0.25 & \\
\hline
\end{tabular}
\end{table*}

\begin{table*}
\caption{Calculated magnetic exchange constants for SrFeF$_5$ and respective Fe -- Fe distances.\label{tab:Js_SrFeF5}}
\centering
\begin{tabular}{ccccccc}
\hline
 & $J_1$ & $J_2$ & $J_3$ & $J_4$ & $J_5$ & $J_6$ \\
\hline
Fe -- Fe, \AA &  3.666 & 3.891 & 4.486 & 4.692 & 5.044 & 5.094 \\
J, meV & 15.5 & 21.4 & 0.05 & 0.38 & 0.34 & -0.91 \\
\hline
\hline
 & $J_7$ & $J_8$ & $J_9$ & $J_{10}$ & $J_{11}$ & $J_{12}$ \\
\hline
Fe -- Fe, \AA & 5.114 & 5.840 & 5.983 & 6.092 & 6.329 & 6.336 \\
J, meV & 1.35 & -0.89 & 0.16 & 2.06 & -1.01 & 1.53 \\
\hline
\end{tabular}
\end{table*}

\section{Magnetic properties}

\begin{figure}
\centering
\includegraphics[width=8.5cm]{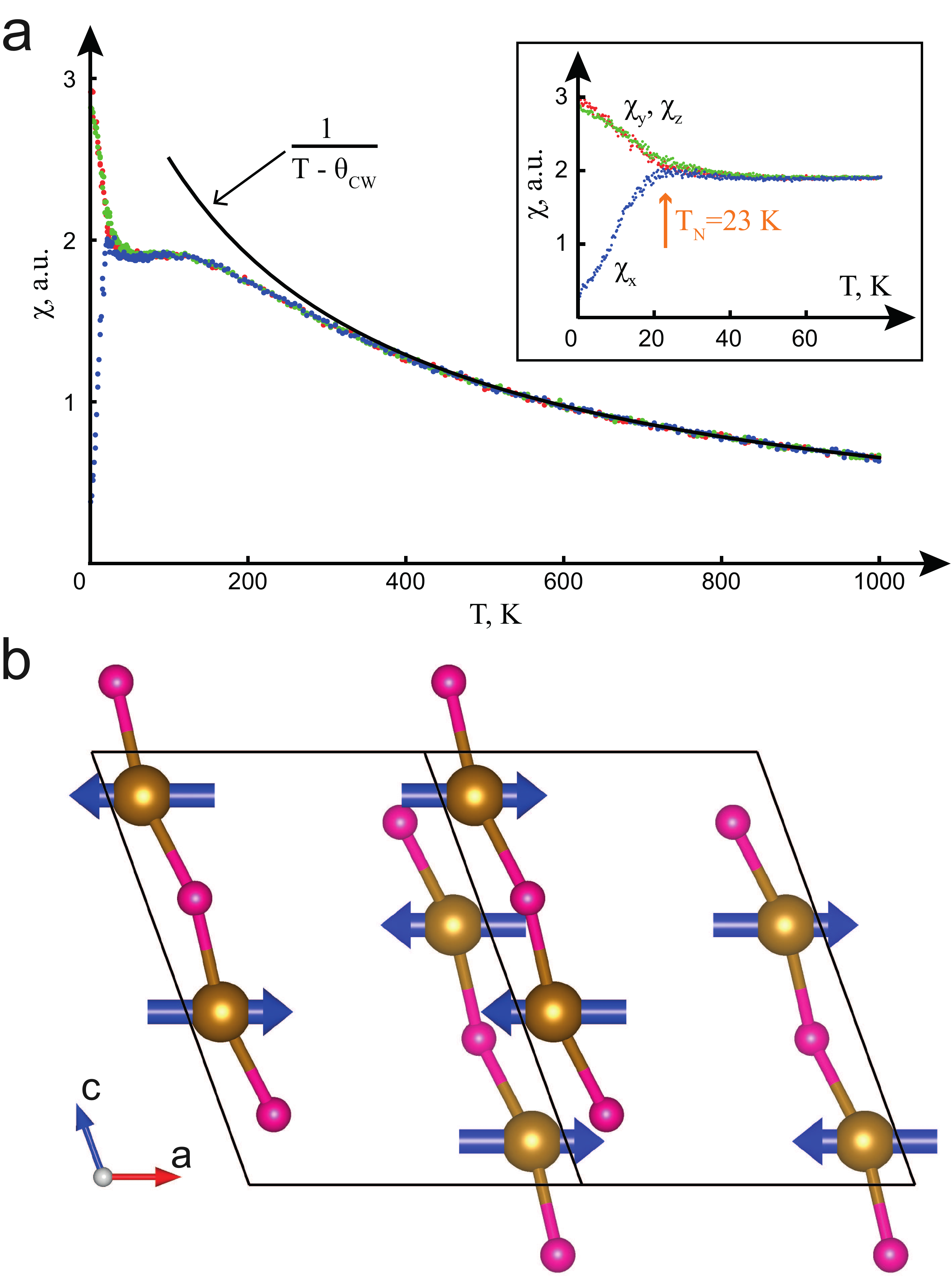}
\caption{\label{fig:CaFeF5_M} (Color online) (a) Magnetic susceptibility $\chi_x$ (blue), $\chi_y$ (green), and $\chi_z$ (red) of CaFeF$_5$ measured along different directions. The inset shows the data below 80~K. (b) Magnetic structure of CaFeF$_5$ at $T=1$~K. Blue arrows indicate the spins of Fe$^{3+}$ ions.}
\end{figure}

The magnetic exchange constants obtained with the help of DFT calculations were used in Monte Carlo simulations. We have not calculated the magnetic anisotropy terms in Eq.~\ref{eq:Hamiltonian} and used tentative values of $D_\alpha$ ($\alpha=x$, $y$, $z$) in our MC studies. In the following we assume that the axes $x$ and $y$ are parallel to the axes $a$ and $b$, respectively, and, the axis $z$ is perpendicular to both the $a$ and $b$ axes. In our MC  calculations of CaFeF$_5$ we used $D_x=-0.1$, $D_y=0.1$, and $D_z=0$~meV. Figure~\ref{fig:CaFeF5_M}(a) shows the temperature dependencies of magnetic susceptibility $\chi$ of CaFeF$_5$ studied along various directions. It can be found that $\chi$ experiences a broad plateau characteristic of 1D magnetic chains structure. The fitting of $\chi$ in the range of temperatures from 500 to 1000~K gives the Curie-Weiss temperature $\theta_{\rm CW}$=-219~K, whereas from the inset in Fig.~\ref{fig:CaFeF5_M}(a) it follows that CaFeF$_5$ experiences an antiferromagnetic phase transition at $T_N=23$~K. Overall the results of calculations are in good agreement with the experimental data on magnetic properties of CaFeF$_5$, which give the Curie-Weiss and Neel temperatures of -202~K and 21~K, respectively~\cite{Dance_1979}. From our point of view the presence of frustrating interchain interactions contribute to the frustration parameter $f=|\theta_{\rm CW}|/T_N\approx9.5$, without which it would have been smaller. The analysis of magnetic structure resulting from MC calculations shown in Fig.~\ref{fig:CaFeF5_M}(b) below $T_N$ reveals that it is characterized by the wave vector $\vec{k}=(1/2,0,0)$. Thus, the magnetic cell is twice the crystallographic one (i.e. doubled along the $a$ axis) and the spins are directed along the easy axis.

\begin{figure}
\centering
\includegraphics[width=8.5cm]{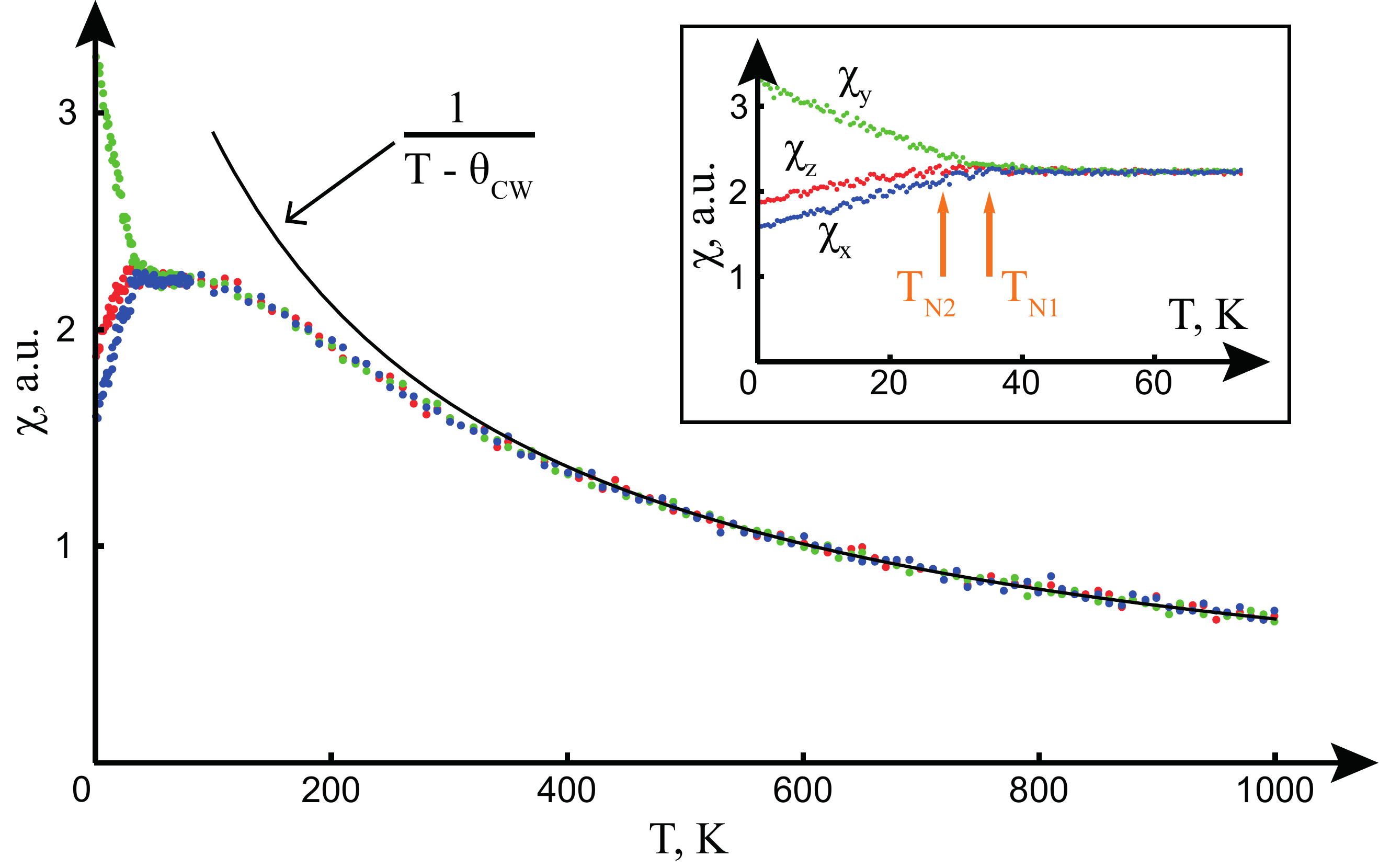}
\caption{\label{fig:SrFeF5_M} (Color online) Magnetic susceptibility $\chi_x$ (blue), $\chi_y$ (green), and $\chi_z$ (red) of SrFeF$_5$ measured along different directions. The inset shows the data below 80~K.}
\end{figure}

Figure~\ref{fig:SrFeF5_M} shows the magnetic susceptibility of SrFeF$_5$ obtained from MC calculations, in which we used the anisotropy constants $D_x=-0.05$, $D_y=0.05$, and $D_z=0$~meV. Similar to CaFeF$_5$ the magnetic susceptibility shows a broad maximum characteristic of a 1D magnetism. The Curie-Weiss temperature obtained by fitting the data in the temperature range 500 -- 1000~K is $\theta_{\rm CW}$=-166~K. In the inset of Fig.~\ref{fig:SrFeF5_M} one can discern two magnetic phase transitions at $T_{N1}=35$ and $T_{N2}=28$~K, corresponding to anomalies in the temperature dependence of $\chi_x$ and $\chi_z$. In SrFeF$_5$, similar to the case of CaFeF$_5$, given a strong tendency for $\ldots$-up-down-up-down-$\ldots$ ordering within the chains, several exchange interactions cannot be simultaneously satisfied (e.g. the pairs $J_5$ and $J_7$, as well as $J_6$ and $J_{11}$), which contribute to both the frustration parameter and to the appearance of incommensurate modulation as described below. 

The analysis of magnetic structure reveals that the first phase transition at $T_{N1}$ is connected with the appearance of sinusoidally modulated structure with spins along the easy axis ($x$ axis in our case), whereas at the second phase transitions at $T_{N2}$ additional component along the $z$ axis appears resulting in cycloidal magnetic structure. The modulation wave vector is $\vec{k}_{inc}\approx(0.22,0,0.32)$. The plane of rotation of spins in the low-temperature magnetic structure is governed by the magnetic anisotropy, which in our model is determined by the tentative choice of the single-ion anisotropy constants $D_\alpha$. Moreover, in our simplified model we assumed the same single-ion anisotropies for both crystallographically different iron positions Fe$_1$ and Fe$_2$, which may differ in general. Thus, the magnetic order in SrFeF$_5$ can in principle be more complex than a cycloidal rotation of spins.

\section{Possible magnetoelectric properties and discussion}

As found in the previous section, the magnetic structure of CaFeF$_5$ below the Neel temperature is characterized by the wave vector $\vec{k}=(1/2,0,0)$, which is the Y-point of the Brillouin zone of the space group P2$_1$/c. The analysis of possible magnetic structures induced by a single exchange multiplet in this point~\cite{Sakhnenko_Exchange_2012}, shows that they cannot induce electric polarization, because they do not break inversion symmetry. This conclusion is also in accord with the absence of modulated magnetic ordering because the irreducible representations (IR) in the Y-point of the Brillouin zone satisfy the Lifshitz criterion~\cite{Sakhnenko_Improper_2010}.

Monte Carlo calculations reveal that the magnetic ordering wave vector in SrFeF$_5$ is $\vec{k}_{inc}\approx(0.22,0,0.32)$. This modulation vector is close to the commensurate vector $\vec{k}_c=(1/4,0,1/3)$, which is the F-point of the Brillouin zone and which is realized in our MC calculations assuming stronger single-ion anisotropy constants $D_\alpha$. The phenomenological treatment of magnetic phase transitions in SrFeF$_5$ can be performed based on the instability with the wave vector $\vec{k}_c$, whereas the incommensurate modulation can be described by Lifshitz invariants. Indeed, the F-point lies in the interior of the Brillouin zone and the two two-dimensional IR's F$_1$ and F$_2$ in this point do not satisfy the Lifshitz criterion. Thus, the crystal symmetry permits the invariants responsible for long-wavelength modulation of magnetic order in the $ac$ plane
\[
a_1\frac{\partial a_2}{\partial x}-a_2\frac{\partial a_1}{\partial x},\qquad a_1\frac{\partial a_2}{\partial z}-a_2\frac{\partial a_1}{\partial z},
\]
\[
b_1\frac{\partial b_2}{\partial x}-b_2\frac{\partial b_1}{\partial x},\qquad b_1\frac{\partial b_2}{\partial z}-b_2\frac{\partial b_1}{\partial z},
\]
where $(a_1,a_2)$ and $(b_1,b_2)$ are magnetic order parameters transforming according to IR's F$_1$ and F$_2$, respectively. Depending on the orientation of spins in the low-temperature magnetic structure of SrFeF$_5$  (e.g. cycloidal, proper screw, or more complex magnetic ordering) at least two magnetic order parameters describe the magnetic structure, which transform either by the same single IR F$_1$ or F$_2$ or by two different IR's F$_1$ and F$_2$. In both cases the magnetoelectric interactions
\[
(a_1b_1+a_2b_2)P_y,
\]
\[
(a_1a'_2-a_2a'_1)P_x, \qquad (a_1a'_2-a_2a'_1)P_z,
\]
\[
(b_1b'_2-b_2b'_1)P_x, \qquad (b_1b'_2-b_2b'_1)P_z,
\]
where the order parameters $(a_1,a_2)$ and $(a'_1,a'_2)$ transform according to F$_1$, while $(b_1,b_2)$ and $(b'_1,b'_2)$ according to F$_2$, should result in appearance of electric polarization $P_\alpha$. Thus, SrFeF$_5$ is predicted to be multiferroic below the magnetic ordering temperature $T_{N2}$, where two magnetic order parameters condense. This picture of magnetically induced electric polarization is similar to that in many other multiferroics~\cite{TerOganessian_2014,Toledano_2009,Toledano_TbMn2O5_2009}. Given the absence of inversion center at the positions of Fe$^{3+}$ ions in SrFeF$_5$ the single-spin effects should contribute to magnetoelectric coupling
~\cite{Sakhnenko_Micro_2012}. These single-ion effects together with the nearest neighbor two-spin couplings along the spin chains are arguably the strongest magnetoelectric interactions in SrFeF$_5$.

Both studied compounds, CaFeF$_5$ and SrFeF$_5$, contain spin chains with strong NN intrachain interactions, weaker NNN intrachain couplings and relatively smaller interchain interactions. In SrFeF$_5$ the interchain couplings are relatively stronger compared with intrachain one, than in CaFeF$_5$, which is probably determined by the helicoidal arrangements of spin chains in SrFeF$_5$ and, thus, shorter interchain distances. In both cases, within the chains the spins order in $\ldots$-up-down-up-down-$\ldots$ fashion, whereas relative ordering of chains is determined by a big number of interchain exchange constants. Our calculations show that\hfill in\hfill CaFeF$_5$\hfill this\hfill results\hfill in\hfill magnetic\hfill ordering\hfill with\hfill\\ $\vec{k}=(1/2,0,0)$, which does not permit magnetoelectric properties, whereas in SrFeF$_5$ a modulated magnetic structure is realized resulting in multiferroic behavior. In both cases the magnetic wave vector is a result of combined action of many exchange parameters. Given possible inaccuracy in determination of exchange constants by DFT calculations the real magnetic structures in CaFeF$_5$ and SrFeF$_5$ can be characterized by wave vectors other, then those found in the present work. However, since the space group P2$_1$/c is non-symmorphic, analysis shows that only the magnetic structures induced by the Y-point instability should not give rise to magnetoelectric properties, whereas all other Brillouin zone boundary points (A, B, C, and Z), which are compatible with the $\ldots$-up-down-up-down-$\ldots$ spin ordering of the chains, contain IR's not satisfying the Lifshitz criterion and, thus, inducing improper electric polarization~\cite{Sakhnenko_Improper_2010}. It has to be noted, that in CaFeF$_5$ the energy of the obtained magnetic structure with $\vec{k}=(1/2,0,0)$ is very close to that of magnetic structure with $\vec{k}=0$, in which the spins of of Fe$^{3+}$ ions located at positions $(x,y,z)$, $(1-x,1-y,1-z)$, $(1-x,1/2+y,1/2-z)$, and $(x,1/2-y,1/2+z)$ are ordered in $(+-+-)$ fashion. Such magnetic ordering breaks inversion symmetry and results in appearance of linear magnetoelectric effect below $T_N$. Neutron diffraction experiments are necessary to clarify the magnetic structures in the studied fluorides.

\section{Conclusions}

We have studied the spin-chain fluorides CaFeF$_5$ and SrFeF$_5$ by means of density functional theory and Monte Carlo calculations, as well as by symmetry analysis. The chains in both compounds are characterized by strong NN exchange couplings and weak NNN interactions. The interchain interactions are much weaker than the NN intrachain couplings, which gives rise to low-dimensional magnetic behavior. Our results indicate that the magnetic structure of CaFeF$_5$ does not result in magnetoelectric effects, whereas SrFeF$_5$ is multiferroic in the low-temperature incommensurate magnetically ordered phase.

\section*{Acknowledgements}

V. O. acknowledges partial financial support from the Science Committee of the Ministry of Education and Science of Armenia (grant SFU-02), ANSEF project condmatth-5212 
as well as the support from the HORIZON 2020 RISE "CoExAN" project 
(GA644076). He also expresses his gratitude to the Institute of Physics, Southern Federal University, Rostov-on-Don for warm hospitality during his visit in 2017.  N.V.T. acknowledges financial support by the Russian Foundation for Basic
Research grant No. 18-52-80028 (BRICS STI Framework Programme).

\appendix
\section{Structural parameters}
\setcounter{table}{0}

Tables~\ref{tab:CaFeF5_structure} and~\ref{tab:SrFeF5_structure} present the lattice parameters and atomic positions of CaFeF$_5$ and SrFeF$_5$ obtained by DFT calculations in comparison with the experimental results taken from literature.

\begin{table}
\caption{Structural parameters of CaFeF$_5$ obtained experimentally ($a$=5.492~\AA, $b$=10.076~\AA, $c$=7.599~\AA, $\beta$=110.02$^\circ$, sp. gr. P2$_1$/c) \cite{Graulich_1998} and by DFT calculations in this work ($a$=5.571~\AA, $b$=10.218~\AA, $c$=7.694~\AA, $\beta$=109.9$^\circ$)\label{tab:CaFeF5_structure}}
\begin{center}
\begin{tabular}{lcccc}      
\hline
Atom &   & $x$ & $y$ & $z$ \\
\hline
Fe & exp.  &   0.10314  &    0.24644  &    0.3996 \\
   & DFT   &   0.10542  &    0.24653  &    0.4001 \\
Ca & exp.  &   0.52690  &   -0.01783  &    0.2540 \\
   & DFT   &   0.52914  &    0.98047  &    0.2548 \\
F$_1$ & exp.  &   0.38640  &    0.11840  &    0.4448 \\
   & DFT      &   0.38841  &    0.11776  &    0.4443 \\
F$_2$ & exp.  &   0.35480  &    0.38740  &    0.4924 \\
   & DFT      &   0.35781  &    0.38768  &    0.4945 \\
F$_3$ & exp.  &   0.12140  &    0.60360  &    0.1878 \\
   & DFT      &   0.12073  &    0.60462  &    0.1880 \\
F$_4$ & exp.  &   0.17220  &    0.86320  &    0.1413 \\
   & DFT      &   0.16907  &    0.86377  &    0.1405 \\
F$_5$ & exp.  &   0.14990  &    0.29350  &    0.1621 \\
   & DFT      &   0.15270  &    0.29490  &    0.1629 \\
\hline
\end{tabular}
\end{center}
\end{table}

\begin{table}
\caption{Structural parameters of SrFeF$_5$ obtained experimentally ($a$=7.062~\AA, $b$=7.289~\AA, $c$=14.704~\AA, $\beta$=95.4$^\circ$, sp. gr. P2$_1$/c) \cite{VonDerMuehll_SrFeF5} and by DFT calculations in this work ($a$=7.127~\AA, $b$=7.395~\AA, $c$=14.891~\AA, $\beta$=95.2$^\circ$)\label{tab:SrFeF5_structure}}
\begin{center}
\begin{tabular}{lcccc}      
\hline
Atom &   & $x$ & $y$ & $z$ \\
\hline
Fe$_1$ & exp.  &  0.5062  &  0.6185  &  0.1614  \\
   & DFT  & 0.5070 & 0.6170 & 0.1611 \\
Fe$_2$ & exp.  &  0.8107  &  0.6832  &  0.3750  \\
   & DFT  & 0.8118 & 0.6846 & 0.3753 \\
Sr$_1$ & exp.  &  0.6896  &  0.1536  &  0.0873  \\
   & DFT  & 0.6847 & 0.1559 & 0.0870 \\
Sr$_2$ & exp.  &  0.0015  &  0.6529  &  0.1374  \\
   & DFT  & 0.0003 & 0.6587 & 0.1385 \\
F$_1$ & exp.  &   0.6977  &  0.4951  &  0.0986  \\
   & DFT  & 0.7001 & 0.4964 & 0.0978 \\
F$_2$ & exp.  &   0.3190  &  0.6978  &  0.0678  \\
   & DFT  & 0.3190 & 0.6920 & 0.0676 \\
F$_3$ & exp.  &   0.3145  &  0.6658  &  0.2414  \\
   & DFT  & 0.3130 & 0.6586 & 0.2417 \\
F$_4$ & exp.  &   0.6797  &  0.5565  &  0.2715  \\
   & DFT  & 0.6831 & 0.5543 & 0.2714 \\
F$_5$ & exp.  &   0.6548  &  0.8301  &  0.1503  \\
   & DFT  & 0.6526 & 0.8293 & 0.1488 \\
F$_6$ & exp.  &   0.9935  &  0.4878  &  0.3911  \\
   & DFT  & 0.9997 & 0.4935 & 0.3890 \\
F$_7$ & exp.  &   0.9680  &  0.8067  &  0.2936  \\
   & DFT  & 0.9669 & 0.8129 & 0.2941 \\
F$_8$ & exp.  &   0.6113  &  0.8693  &  0.3540  \\
   & DFT  & 0.6036 & 0.8617 & 0.3560 \\
F$_9$ & exp.  &   0.6531  &  0.5802  &  0.4594  \\
   & DFT  & 0.6545 & 0.5734 & 0.4584 \\
F$_{10}$ & exp.  &  0.9559  &  0.8206  &  0.4657  \\
   & DFT  & 0.9561 & 0.8205 & 0.4677 \\
\hline
\end{tabular}
\end{center}
\end{table}

\end{document}